\documentclass[fleqn,usenatbib]{mnras}

\usepackage{newtxtext,newtxmath}

\usepackage[T1]{fontenc}

\DeclareRobustCommand{\VAN}[3]{#2}
\let\VANthebibliography\thebibliography
\def\thebibliography{\DeclareRobustCommand{\VAN}[3]{##3}\VANthebibliography}

\usepackage{graphicx}	
\usepackage{amsmath}	
\usepackage[lofdepth,lotdepth]{subfig}

\DeclareUnicodeCharacter{2212}{-}

\title[Bispectrum of QPOs from GRS~1915+105]{Using the bispectrum to probe radio X-ray correlations in GRS~1915+105 }

\author[Arur \& Maccarone.]{
K. Arur,$^{1}$\thanks{E-mail: karur3@gatech.edu}
T. J. Maccarone,$^{2}$
\\
$^{1}$Center for Relativistic Astrophysics, School of Physics, Georgia Institute of Technology, 837 State Street, Atlanta, GA 30332-0430, USA\\
$^{2}$Department of Physics and Astronomy, Texas Tech University, Lubbock, TX, 79409-1051,USA\\
}

\date{Accepted XXX. Received YYY; in original form ZZZ}

\pubyear{2021}

\begin{document}
\label{firstpage}
\pagerange{\pageref{firstpage}--\pageref{lastpage}}
\maketitle

\begin{abstract}
We present the results of bicoherence analysis on observations of GRS~1915+105 that exhibit quasi-periodic oscillations (QPOs). The bicoherence is a higher order statistic that can be used to probe the relation between the phases of a triplet of Fourier frequencies. Despite showing very similar power spectra, the observations exhibit different patterns in their bicoherence, indicating that the QPOs are phase coupled to the noise in different ways. We show that the bicoherence pattern exhibited correlates with the frequency of the QPO, the hardness ratio, as well as the radio properties of the source. In particular, we find that the nature of phase coupling between the QPO and the high and low frequency broadband components is different between radio quiet, radio plateau and radio steep conditions. We also investigate the phase lag behaviour of observations with QPO frequency above 2Hz that show different bicoherence patterns, and find statistically significant differences between them, indicating a change in the underlying physical mechanism. Finally, we present a scenario whereby the cooling of the jet electrons by soft photons from the accretion disc could explain the observed correlations between the bicoherence and radio properties.
\end{abstract}

\begin{keywords}
X-rays: binaries -- accretion, accretion discs -- methods: statistical
\end{keywords}



\section{Introduction}

GRS~1915+105 is a low mass black hole X-ray binary that was discovered in 1992 by the GRANAT satellite \citep{Castro-Tirado1992} and has since become one of the best monitored X-ray sources in the sky. The X- ray light curves of GRS~1915+105 have been classified into at least 14 variability classes based on their colour and timing properties \citep{Belloni2000,Klein-Wolt2002a,Hannikainen2003,Hannikainen2005}. These classes, that repeat on timescales of months to years, are thought to be driven by global instabilities in the accretion disc. However, these classes are not readily explained by standard accretion theory, with only one other source (IGR J17091-3624) exhibiting similar types of variability \citep{Altamirano2011}. The rich dataset available as a result of extensive monitoring of this system has also made the classification of these distinctive states well suited for methods involving machine learning algorithms \citep{Vitanyi2013,Huppenkothen2017}. These classes are interpreted as transitions between 3 spectral states A, B and C (defined based on the flux and hardness ratios of the source at the time of observation) The hard C state is characterised by low flux, while States A and B are much softer, and are characterised by low and high flux respectively.  All three states show similarities to the Very High State seen from other black hole X-ray binaries \citep{Reig2003}.

The radio properties of GRS~1915+105 are also remarkable, showing superluminal jets \cite{Mirabel1994}. Continuous radio monitoring helped with the identification of two radio bright states: the plateau state with optically thick, flat radio spectra and the flaring state with optically thin spectra using the Green Bank Interferometer (GBI) at 2.3 and 8.3 GHz \citep{Foster1996}. The flaring states were found to correspond to relativistic ejections, with the energies associated with particle acceleration and bulk motion of the jet being a significant fraction of the accretion power \citep{Mirabel1994,Fender1999}. On the other hand, the radio plateau state is associated with a steady radio jet, which can be explained by synchrotron self-absorption. For a detailed overview of the radio properties of GRS~1915+105, see \cite{Fender2004}.

Quasi-Periodic Oscillations (QPOs) have been detected from the X-ray light curves of GRS~1915 in several classes. These QPOs occur in a range of frequencies, including very low frequency (< 0.1 Hz), high amplitude QPOs, low frequency QPOs in the 0.1-10Hz range, a stable 67Hz QPO that is sometimes seen, and high frequency ($\sim$ 170 Hz) QPOs \citep{Morgan1997,Belloni2006}. In this paper, we focus on the low frequency QPOs seen between 0.1 and 10Hz, thought to be analogous to type-C QPOs from other black hole X-ray binaries \citep{Casella2005}.

These low frequency QPOs in GRS~1915+105 have been extensively studied, and shown to exhibit complex behaviour. The QPO frequency from GRS~1915+105 was found to be energy dependent, with QPOs with a centroid frequency below 2Hz showing frequency that decreases with photon energy, while the QPO frequency increases with energy above 2Hz \citep{Yan2012}. \cite{Reig2000a} found that the phase lag at the QPO fundamental frequency evolves smoothly as a function of the QPO frequency, and switches from a positive (hard) lag to a negative (soft) lag at ~2Hz. QPOs from GRS~1915+105 also show evolution on sub second timescales, with the QPO decohering on timescales of 5-10 QPO cycles, before resetting \citep{Eijnden2016}. \cite{Yan2013} found that QPOs from GRS~1915 could be divided into two branches based on the X-ray luminosity and hardness ratio of the observations and found that these branches correlate with the radio properties of the source. They also found that the fractional rms amplitude of the QPO increases with frequency when the centroid frequency was below 2Hz. However, for frequencies above 2Hz, the QPO amplitude decreases with frequency. 

While the QPO phenomenon has been well studied emprically, the physical mechanism that produces the QPO is still under debate. Multiple models have been proposed, and these models can be broadly divided into two categories: intrinsic, where changes intrinsic to the system such as the pressure or accretion rate modulate the X-ray flux, and geometric models, where the QPO is caused by a quasi-periodic change in the geometry of the system as viewed by an observer. For a detailed review on the observational properties of QPOs and the proposed models, we refer the reader to \cite{Ingram2019}. 

The inclination dependence of QPO properties such as the QPO amplitude \citep{Motta2015, Heil2015}, phase lag behaviour \citep{VandenEijnden2017} and non linear variability \citep{Arur2020}; as well as the modulation of the iron line \citep{Ingram2015} are suggestive of a geometric origin for the QPO. One explanation for this geometric origin is that the QPO is caused by the Lense-Thirring precession of the Comptonizing corona \citep{Stella1998, Ingram2009}, which is thought to produce the power law component of the X-ray spectrum via inverse Compton scattering of photons. However, models that that have both intrinsic and geometric components, such as the accretion ejection instability model \citep{Tagger1999a,Varniere2017} could also explain such inclination dependence. 

Correlations between radio and X-ray variabilty have also been studied to probe the connection between the accretion disc and radio jet in GRS~1915+105. \cite{Klein-Wolt2002a} found a relation between radio oscillation events and series of long State C intervals, with each State C interval producing a radio flare. Additionally, \cite{Muno2001} examined the evolution of QPO properties during periods of faint and bright radio emission, and found that as the radio emission becomes brighter, the QPO frequency decreases. However, the link between the base of the jet and the Componising corona is poorly understood. Additionally, there is no consensus on the origin of the seed photons for the inverse Compton process that is thought to be the origin of the power law component of the X-ray spectrum. Whether the dominant source of these seed photons arise from the thermal accretion disc \citep{Thorne1975}, synchrotron radiation \citep{Narayan1994}, or switches between the two with a change in the accretion state \citep{Skipper2013} is presently under debate.

A powerful way to break degeneracies between different models that can reproduce the power spectra is to use higher order timing analysis techniques, as demonstrated by \cite{Maccarone2005} for the case of high frequency QPOs. The bispectrum, which is the Fourier domain equivalent to the three point correlation function, is one such method. The closely related bicoherence is a measure of the strength of the coupling among the phases of the different Fourier frequency components. Study of the  bicoherence of observations from GRS 1915 that exhibit QPOs \citep{Maccarone2011} revealed that the QPO and broadband noise frequencies are phase coupled, indicating that the two processes are not generated independently. It was also found that despite having very similar power spectra, the bispectra show different behaviour which were phenomenologically classified into three patterns of variability (see Section~\ref{sec:bicoherence_patterns}). \cite{Maccarone2011} also found that the patterns seen in the bicoherence of GRS~1915 QPOs varied based on the radio properties of the source. 

In this paper, we analyse a sample of $\sim$600 observations of GRS~1915+105 to perform a detailed examination on how the properties of the X-ray lightcurve such as count rate, hardness ratio and QPO frequency affect the properties of the bispectrum. We focus on the low frequency QPOs observed during the C state ($\chi$ class), which are considered to be analogous to Type-C QPOs observed from other black hole binaries. 

\section{The Bispectrum}

In this work, we focus on the computation of the bispectrum \citep{Tukey1953,Hasselmann1963}, and the closely related bicoherence. The bispectrum, which is the Fourier domain equivalent of the 3-point correlation function, is given by:

\begin{equation}
    B(k,l) =\frac{1}{K} \sum_{i=0}^{K-1}X_i(k)X_i(l)X^*_i(k+l)
	\label{eq:bispectrum}
\end{equation}

where $X_i(f)$ is the Fourier transform of the $i$th segment of the time series at frequency $f$, and $X^*_i(f)$ is the complex conjugate of $X_i(f)$ \citep{Fackrell1996}  

A related term is the bicoherence, which is given by: 

\begin{equation}
    b^2 =\frac{\left|\sum X_i(k)X_i(l)X^*_i(k+l) \right|^2}
    {\sum|X_i(k)X_i(l)|^2 \sum |X_i(k+l)|^2} 
	\label{eq:bicoherence}
\end{equation}

using the normalization proposed by \cite{Kim1979}. Here, $b^2$ measures the fraction of power at the frequency $k+l$ due to the coupling of the three frequencies. The value of the bicoherence lies between 0 and 1, with 0 indicating that no phase coupling is present among the frequencies, and 1 indicating total phase coupling. We note that in this paper we focus our analysis on low frequency QPOs where the frequencies are not dominated by effects of Poisson noise or dead time effects. A bias of 1/K \citep{Maccarone2002} is subtracted from all bicoherence measurements presented in this paper.

The phase of the bispectrum, known as the biphase, contains important information about the shape of the underlying lightcurve such as the reversibility of the time series in a statistical sense and the skewness about the mean of the flux (see \cite{Maccarone2013} for a details on the biphase). The biphase can thus be used to reproduce the underlying QPO waveform \citep{Arur2019}. The biphase behaviour of QPOs from GRS~1915 will be analysed in future work. 

\subsection{Data Reduction and Analysis}

For our analysis, we consider Rossi X-ray Timing Explorer Proportional Counter Array (PCA) observations of GRS~1915+105 where a LFQPO is present, based on the list presented in \cite{Yan2013} as well as the observations analyzed by \cite{Zhang2020} (see Table~\ref{tab:observations}) which occurred when the source was in the $\chi$ class.

\begin{table*}
  \renewcommand{\thetable}{\arabic{table}a}
  \centering
  \caption{Table of observations used for the analysis in this paper, along with the QPO frequency and Q value, count rate, hardness ratio (13-30keV/2-13keV) and bicoherence pattern of each observation. The full table can be found in the online version of the paper.}
  \label{tab:observations}
  \begin{tabular}{lccccccr} 
		\hline
		No. & OBSID & Obsdate & Frequency & Q & Count rate & Hardness & Bicoherence\\
		& & (MJD) & (Hz) & & (cts/s/PCU) & Ratio & Pattern \\
		\hline
		1 & 10258-01-02-00 & 50293.77 & 0.661 & 5.14 & 1710 & 0.088 & Web \\
		2 & 10258-01-03-00 & 50301.30 & 1.496 & 3.40 & 1725 & 0.073 & Web \\ 
		3 & 10258-01-04-00 & 50309.51 & 3.134 & 6.41 & 1899 & 0.061 & Hypotenuse \\
		\hline
	\end{tabular}
\end{table*}
\begin{table*}
  \addtocounter{table}{-1}
  \renewcommand{\thetable}{\arabic{table}b}
  \caption{Table of bicoherence values for observations used for the analysis in this paper. The description of the columns is as follows: Diagonal - mean value of bicoherence of along the region where $f_1 + f_2 = f_{QPO}$, Cross 1 - mean value of bicoherence of along the region where $f_{QPO} < f_1 < $2$f_{QPO}$, Cross 2 - mean value of bicoherence of along the region where 2$f_{QPO} < f_1 < $4$f_{QPO}$, Background - mean value of bicoherence of in a region where no interaction from a QPO is expected. The columns labelled `Uncert' give the 1$\sigma$ uncertainty on the corresponding bicoherence value. Note that the mean bicoherence value in the `background' region can be negative due to bias subtraction. The full table can be found in the online version of the paper.}
  \label{tab:bc_vals}
  \begin{tabular}[c]{cccccccccc}
		\hline
	No. & OBSID & Diagonal & Diagonal & Cross 1 & Cross 1 & Cross 2 & Cross 2 & Background & Background \\
	    &	    &        &    Uncert  &          &  Uncert  &       & Uncert  &           & Uncert \\
		\hline
    1 & 10258-01-02-00 & 0.0343 & 0.0026 & 0.0392 & 0.0126 & 0.0632 & 0.0074 & -0.0006 & 0.0002 \\
    2 & 10258-01-03-00 & 0.0078 & 0.0009 & 0.0071 & 0.0012 & 0.0047 & 0.0004 & 0.0007 & 0.0005 \\
    3 & 10258-01-04-00 & 0.0021 & 0.0002 & 0.0026 & 0.0003 & 0.001 & 0.0001 & 0.0003 & 0.0004 \\
		\hline
	\end{tabular}
\end{table*}

Light curves were extracted from binned mode data using the standard FTOOLs package, with a time resolution of 1/128s (Nyquist frequency of 64 Hz) and background subtracted using pcabackest. 16 second long segments were used to produce the averaged power density spectrum (PDS) that were normalized according to \cite{Leahy1983}. This time resolution and segment length were also used to calculate the bicoherence according to Equation~\ref{eq:bicoherence}.  

The PDS were logarithmically re-binned (re-bin factor of 1.03) converted to square fractional rms and loaded into XSPEC \citep{Arnaud1996} with a one-to-one correspondence between energy and frequency. The PDS were then fit with a power law and multiple lorentzian components to model the noise and LFQPO features (including higher or sub-harmonic features). These fits were used to determine the QPO fundamental frequency and Q value\footnote{defined as $f_{QPO}$/FWHM, where $f_{QPO}$ is the centroid frequency of the Lorentzian used to fit QPO, and FWHM is its full width at half maximum}. In some observations, poor fits were obtained due to frequency drift of the QPO. In such cases, a portion of the observation where the QPO frequency is stable was used to obtain better fits.

The bicoherence patterns of the observations were classified into `hypotenuse', `cross' and `web' following \cite{Maccarone2011}. Examples of each pattern are shown in Figure~\ref{fig:hypotenuse}, \ref{fig:cross} and \ref{fig:web} respectively.

For each observation, the mean value of the bicoherence along the diagonal region (i.e where $f_1+f_2$ = $f_{QPO} \pm$ FWHM/2) and along the vertical/horizontal region (i.e where $f_1 = f_{QPO} \pm$ FWHM/2) was calculated after bias subtraction. In all cases, the regions where the coupling between the QPO and the (sub)harmonic would dominate were excluded in this calculation. This value was then compared to the mean and variance of the bicoherence from a region without any QPO interactions, where the Poisson noise dominates. If the bicoherence in the regions of interest was at least 5$\sigma$ greater than in the Poisson dominated region, the bicoherence in the region of interest was classified to be significant. 

If the bicoherence in both the diagonal and vertical/horizontal regions was significant, the observation was classified as `web'. In the cases where either only the diagonal or vertical/horizontal regions was significant, these observations were classified as `hypotenuse' or `cross' respectively. If neither component was detected, the observation was classified as `None'. The bicoherence values in these different regions for each observation are included (see \ref{tab:bc_vals}) in the supplementary data table that can be found in the online version of this paper.  We note that in some observations, no significant bicoherence is detected due to either a short observation length, low count rate, low number of active detectors, or a combination of these factors. Observations where the count rates were lower than 900cts/s/PCU were excluded as these observations do not have sufficient statistics to reliably estimate the bicoherence. In a few very long observations, features in the bicoherence plot of the entire observation were smeared out due to QPO drift. In such cases, the pattern from a portion of the observation was used for the classification. No change in the bicoherence pattern was observed during the course of a single observation. 

In order to calculate the hardness ratio, light curves from the $\sim$2-13keV and $\sim$13-30keV energy bands were extracted from Standard2 mode data. Due to changes in the PCA gain, the exact energy channels used differ slightly based on the epoch of the observation. Details of the energy channels used are listed in Table~\ref{tab:energy}. To correct for any changes in the effective area of the detectors at different different epochs, a correction to the hardness ratio was made by normalising the measured flux with the corresponding flux from the Crab nebula using the channels listed in Table~\ref{tab:energy} for each epoch. We find that the correction is small, and does not significantly affect the results discussed below.

In order to produce a frequency-dependent phase lag spectrum (lag-frequency spectrum) for each observation between the hard and soft energy bands, the Cross-Correlation Function (CCF) was calculated, which is given by $C(f) = F_1^*(f)*F_2(f)$, where $F_1(f)$ and $F_2(f)$ are the Fourier transformations of the light curves of the two energy bands. The argument of the CCF is then the phase lag ($\Delta \phi$) \citep{Vaughan1997}. In order to determine the phase lag at the QPO fundamental frequency, the phase lags were averaged over the width of each Fourier component, around its centroid frequency, in the frequency range of $\nu \pm$ FWHM/2. The phase lags were calculated between the 2-5.7 keV and 5.7-15 keV energy bands\footnote{the exact energy bands differ slightly due to gain corrections} following \cite{Zhang2020}. They also find that the dead-time driven cross-talk effects for these observations were found to be negligible and thus no corrections were made for this. 

The error on the phase lags was estimated using the errors on the real and imaginary components on the phase lags \citep{Cui1997}

\begin{equation}
    \delta \Delta \phi = |0.5 {\rm sin}(2\Delta \phi)| \left ( \left |\frac{\delta \langle R \rangle}{\langle R \rangle} \right |+ \left |\frac{\delta \langle I \rangle}{\langle I \rangle} \right | \right )  
\end{equation}

where the angled brackets represent ensemble averaging.

\subsection{Bicoherence Patterns}
\label{sec:bicoherence_patterns}

\subsubsection{Hypotenuse}
\begin{figure}
	\includegraphics[width=\columnwidth]{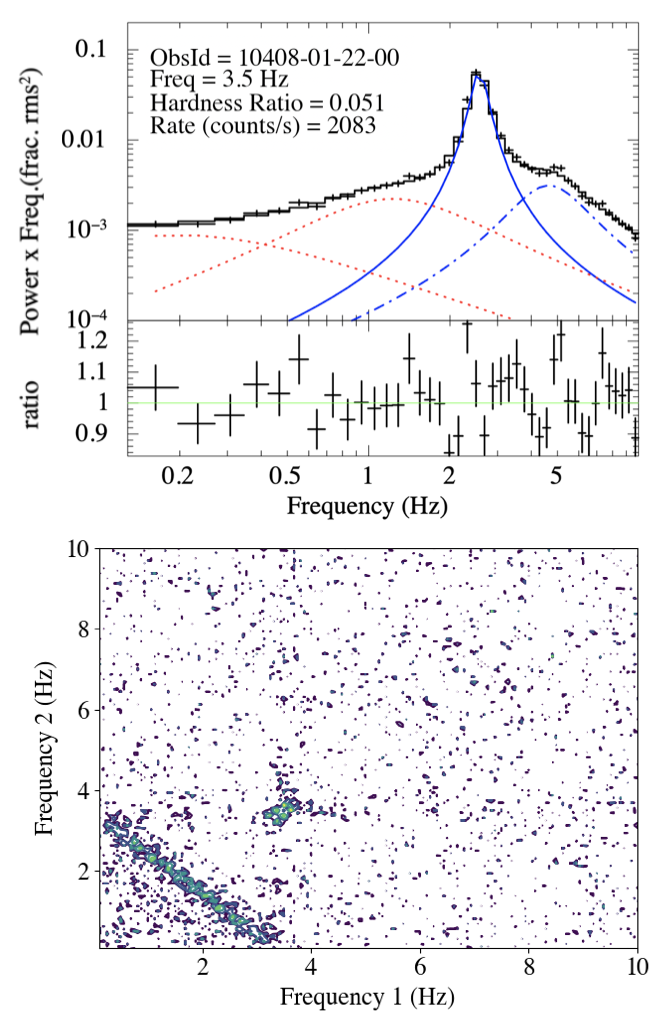}
    \caption{The power spectrum along with the fitted components and ratio of the total model to the data. (Top and Middle panel). The solid blue line indicates the QPO fundamental, the dot-dash blue line(s) indicate the (sub)harmonics and the dotted red line(s) indicate the broadband noise components. Poisson noise has been subtracted from the power spectrum. The bicoherence plot (Bottom panel) shows an example of the 'hypotenuse' pattern from observation 10408-01-22-00. The diagonal region on the bottom left of the plot indicates interaction between the low frequency noise and the QPO fundamental. The colour scheme of log$b^2$ is as follows: purple:-2.0, blue:-1.75, green:-1.50, yellow:-1.25}
    \label{fig:hypotenuse}
\end{figure}

The `hypotenuse' pattern is seen when high bicoherence is found  in the diagonal region where the two frequency components add up to the frequency of the QPO (see Fig.~\ref{fig:hypotenuse} for an example). This shows that the phases of the QPO fundamental and the low frequency ($f < f_{QPO}$) broadband noise are coupled.

This pattern also shows a region of high bicoherence where both $f_1$ and $f_2$ are equal to $f_{QPO}$. This indicates coupling between the fundamental and second harmonic components.

\subsubsection{Cross}

\begin{figure}
	\includegraphics[width=\columnwidth]{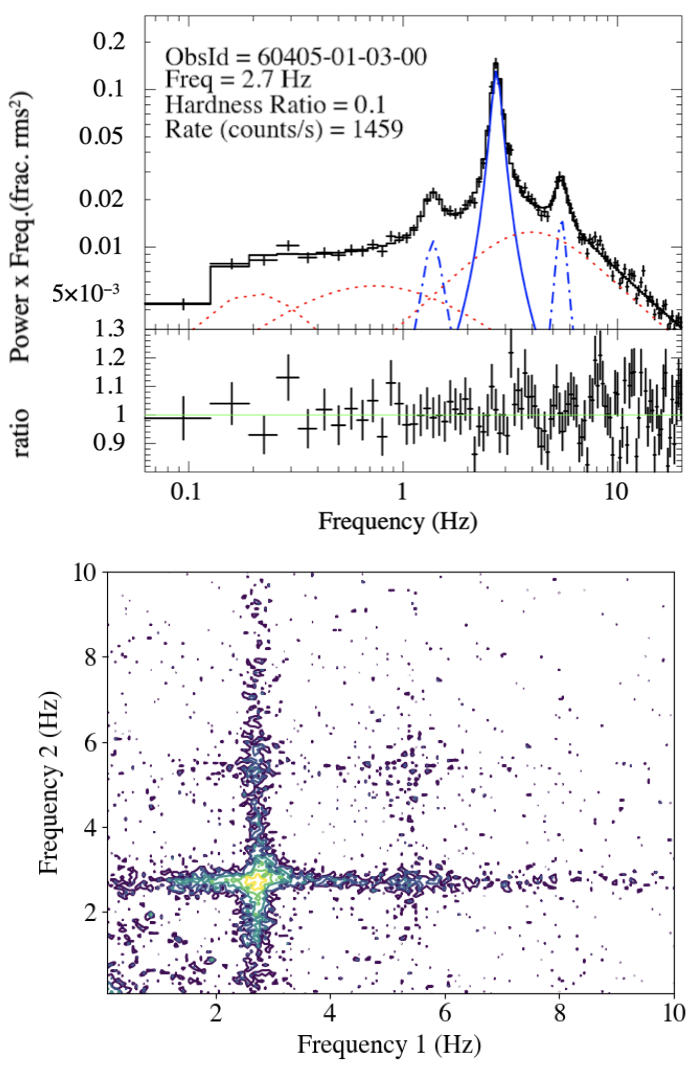}
    \caption{The power spectrum along with the fitted components and ratio of the total model to the data. (Top and Middle panel). The solid blue line indicates the QPO fundamental, the dot-dash blue line(s) indicate the (sub)harmonics and the dotted red line(s) indicate the broadband noise components. Poisson noise has been subtracted from the power spectrum. The bicoherence plot (Bottom panel) shows an example of the 'cross' pattern from observation 60405-01-03-00. The the vertical and horizontal features indicate interaction between the high frequency noise and the QPO fundamental. The colour scheme of log$b^2$ is as follows: purple:-2.0, blue:-1.75, green:-1.50, yellow:-1.25}
    \label{fig:cross}
\end{figure}

In the `cross' pattern, high bicoherence is seen for frequency pairs where one frequency is that of the QPO, and the other frequency can be of any value. This leads to the prominent vertical and horizontal streaks as seen in Fig.~\ref{fig:cross}. This shows that the phases of the QPO fundamental and the high frequency ($f > f_{QPO}$) broadband noise are coupled.

Like in the case of the hypotenuse, high bicoherence is seen in the region of $f_1$ = $f_2$ = $f_{QPO}$. Additionally, high bicoherence is also typically seen in the region where $f_1$ = $f_{QPO}$ and $f_2$ = $2f_{QPO}$ and in the region where $f_1$ = $f_2$ = $2f_{QPO}$. These features indicate phase coupling among the QPO fundamental and second harmonic to produce the third harmonic; and phase coupling of the second harmonic with itself to produce the fourth harmonic respectively.

\subsubsection{Web}

\begin{figure}
	\includegraphics[width=\columnwidth]{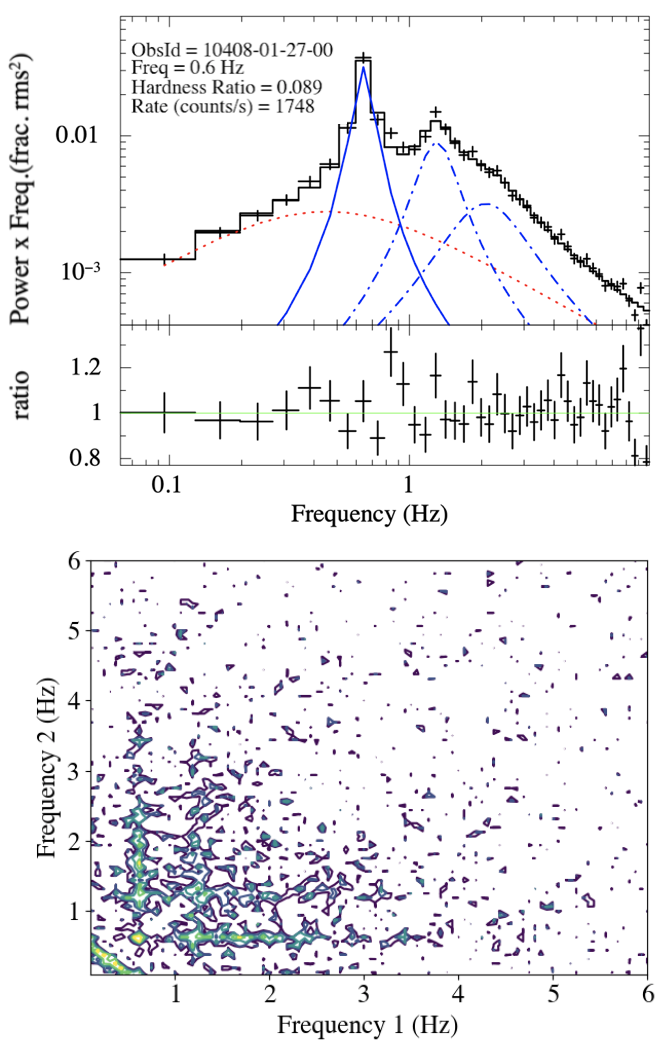}
    \caption{The power spectrum along with the fitted components and ratio of the total model to the data. (Top and Middle panel). The solid blue line indicates the QPO fundamental, the dot-dash blue line(s) indicate the (sub)harmonics and the dotted red line(s) indicate the broadband noise components. Poisson noise has been subtracted from the power spectrum. The bicoherence plot (Bottom panel) shows an example of the 'web' pattern from observation 10408-01-27-00. This pattern is a hybrid, showing features from both the hypotenuse and cross patterns. The colour scheme of log$b^2$ is as follows: purple:-2.0, blue:-1.75, green:-1.50, yellow:-1.25}
    \label{fig:web}
\end{figure}

The `web' pattern is a hybrid class (see Fig.~\ref{fig:web} for an example), where both the diagonal feature of the `hypotenuse' and the vertical and horizontal streaks from the `cross' pattern are seen. When this pattern is observed, the phase at the QPO fundamental frequency is coupled to that of both the high and low frequency broadband noise. Phase coupling is also present between the QPO fundamental and higher harmonics. This pattern is almost always seen in observations where the QPO frequency is low (< $\sim$2 Hz). However, this pattern is also seen in GRS~1915+105 when higher QPOs are present. This is discussed in further detail in Section~\ref{sec:freq_dist}

\begin{table}
	\centering
	\caption{Table of the energy ranges used to calculate the hardness ratio.}
	\label{tab:energy}
	\begin{tabular}{lcccr} 
		\hline
		Epoch & Start & End & Channel & Energy Range\\
		&(MJD)& (MJD) &  Range & (keV) \\
		\hline
		3 & 50188 & 51259 & 0-35 & 1.94 - 12.99\\
		3 & 50188 & 51259 & 36-81 & 13.36 - 30.02\\
		4 & 51259 & 51677 & 0-30 & 2.13 - 13.06\\
		4 & 51259 & 51677 & 31-69 & 13.48 - 29.97\\
		5 & 51677 & 55931 & 0-30 & 2.06 - 12.69 \\
		5 & 51677 & 55931 & 31-71 & 13.11 - 29.97 \\
		\hline
	\end{tabular}
\end{table}

\section{Results}

\subsection{Frequency distribution of bicoherence}
\label{sec:freq_dist}

\begin{figure}
	\includegraphics[width=\columnwidth]{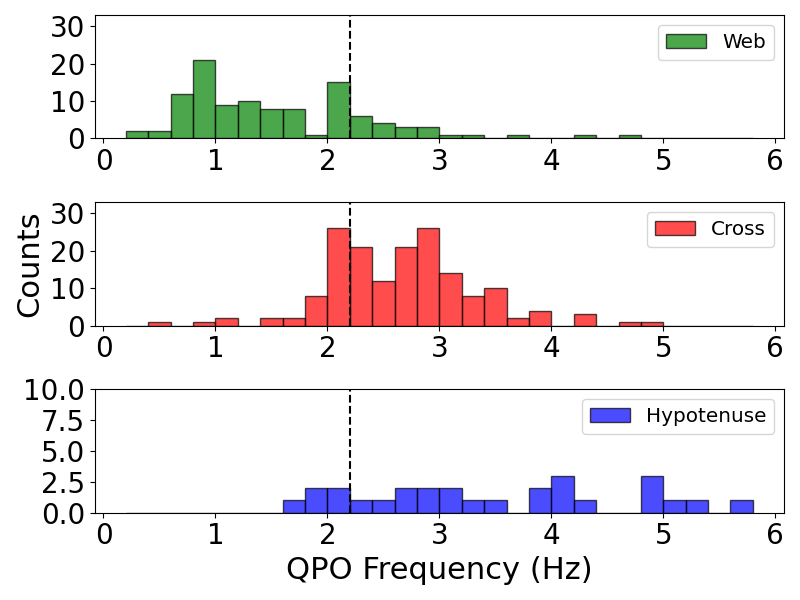}
    \caption{The distribution of the bicoherence patterns as a function of QPO frequency. The dashed vertical line indicates a QPO frequency of 2.2Hz.}
    \label{fig:pattern_freq}
\end{figure}

The distribution of the bicoherence patterns as a function of the QPO frequency is shown in Figure~\ref{fig:pattern_freq}.When the QPO frequency is below $\sim$2 Hz, the `web' pattern is predominantly seen, in agreement with previous studies of the bicoherence of black hole X-ray binaries \citep{Maccarone2011, Arur2019, Arur2020}. Around the 2-2.2Hz frequency, an overlap is seen between the `web' and the `cross' patterns. It has been shown that the bicoherence pattern shows a gradual change from `web' to either a `cross' or a `hypotenuse' pattern, based on the source inclination. As a high inclination (edge-on) source, this overlap between `web' and `cross' pattern in GRS~1915+105 is consistent with the behaviour seen from other high inclination black hole X-ray binaries \citep{Arur2020}.)

\begin{figure}
	\includegraphics[width=\columnwidth]{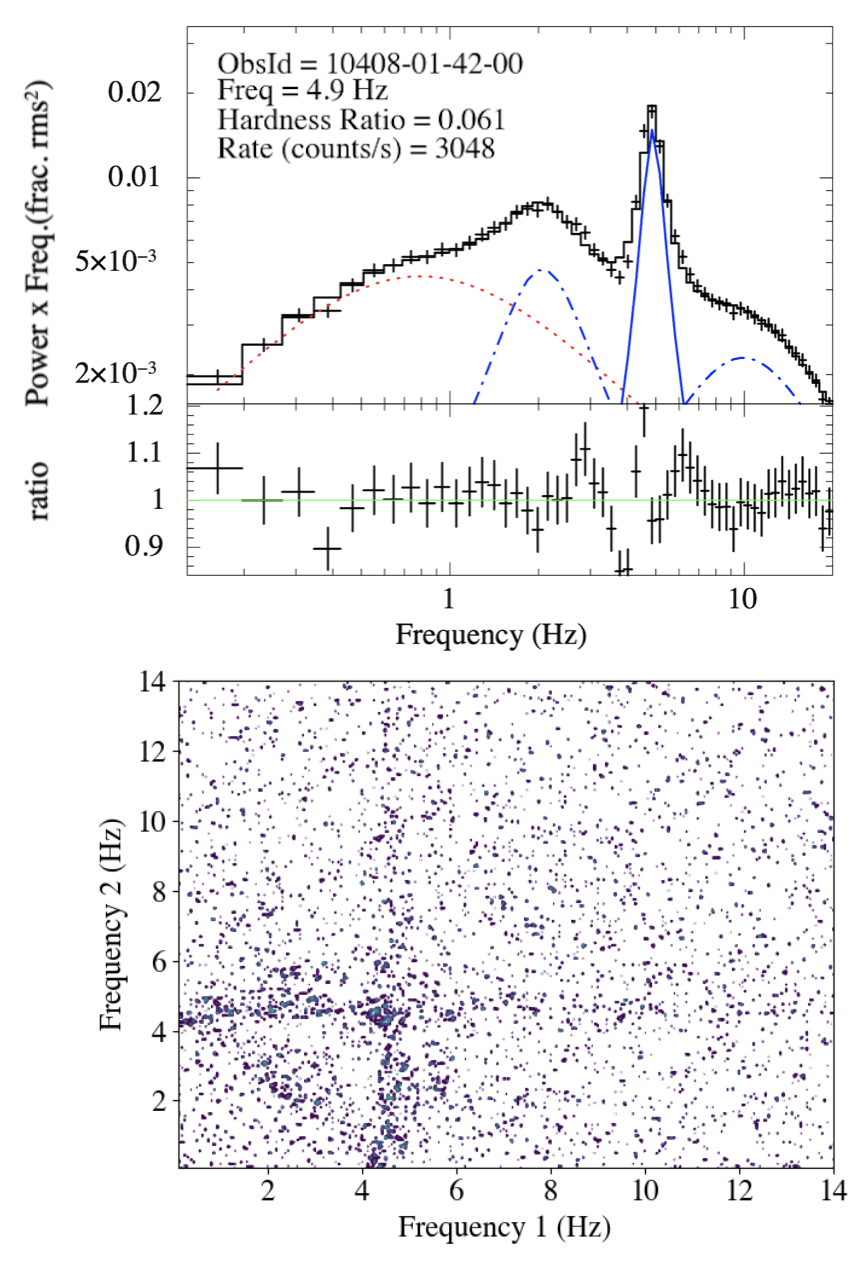}
    \caption{The power spectrum along with the fitted components and ratio of the total model to the data. (Top and Middle panel). The solid blue line indicates the QPO fundamental, the dot-dash blue line(s) indicate the (sub)harmonics and the dotted red line(s) indicate the broadband noise components. Poisson noise has been subtracted from the power spectrum. The bicoherence plot (Bottom panel) shows an example of the 'web' pattern from observation 10408-01-42-00, which has a frequency >2Hz. The diagonal region on the bottom left of the plot is likely to be due to the presence of a broad sub-harmonic feature, visible in the power spectrum. The colour scheme of log$b^2$ is as follows: purple:-2.0, blue:-1.75, green:-1.50, yellow:-1.25}
    \label{fig:web_hf}
\end{figure}

When the QPO frequency is below 2Hz, the `web' pattern is predominantly seen, this pattern is also sometimes seen at higher QPO frequencies. However, in these cases (particularly when the QPO frequency is greater than $\sim$ 3Hz, the diagonal feature on the lower left corner does not fully extend down to the lowest frequencies (See Fig~\ref{fig:web_hf} for an example). It can be seen from the power spectrum of these observations that a broad feature can be seen at half the frequency of the QPO fundamental. Thus, it is likely that the presence of an extended sub-harmonic like feature in the bicoherence mimics the appearance of a `web' pattern in observations that would otherwise be classified as `cross'. We also note that \cite{Reig2000a} found that in the phase lag spectra of observations where the QPO frequency is greater than 3.5Hz, a dip is seen at low frequencies. This could further indicate that this reappearance is due to the presence of a broad sub-harmonic feature.

A number of the observations where no discernible pattern is observed (labelled `None') have QPOs with frequencies above 4Hz, and the count rate is high (>2000 counts/s/PCU). As the bicoherence is calculated by averaging over multiple segments of an observations, in the cases where source is not stationary over the length of the entire observation, reliable estimates of the bicoherence cannot be obtained.

\subsection{Hardness Intensity Diagram}
\label{sec:HID}

\begin{figure}
	\includegraphics[width=\columnwidth]{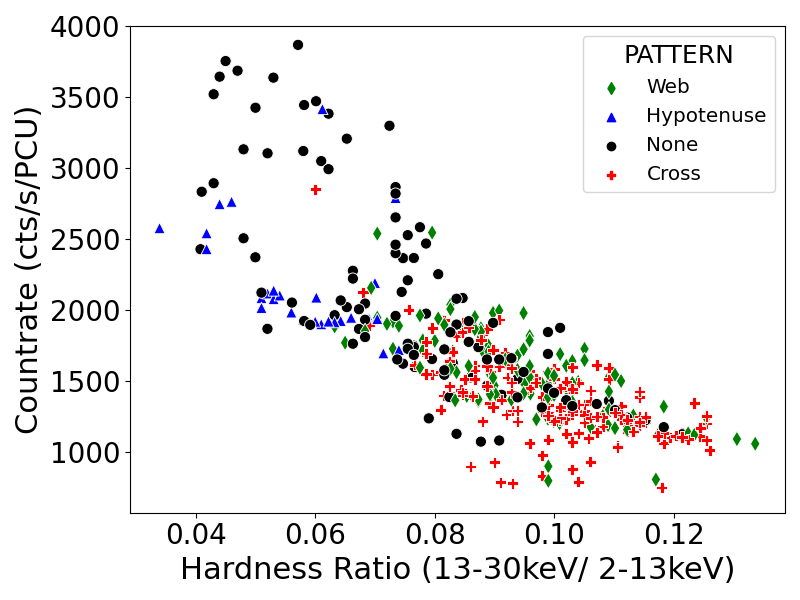}
    \caption{The bicoherence patterns of QPOs plotted on a Hardness-Intensity diagram. The coloured markers indicate the bicoherence pattern seen in each observation.}
    \label{fig:HID}
\end{figure}

The distribution of bicoherence patterns on the hardness-intensity diagram is shown in Fig.~\ref{fig:HID}. It can be seen that when the source is hard (HR $\approx$ 0.09-0.1), at lower count rates ($\sim$1400 cts/s/PCU) the `cross' pattern is seen. However, at similar hardness ratios  when the count rate is slightly higher ($\sim$1600 cts/s/PCU), the `web' pattern is seen.

At high count rates ($\sim$2200 cts/s/PCU), and when the hardness ratio is lower (i.e softer, HR $\approx$ 0.06), the `hypotenuse' pattern is observed. As outlined in Section~\ref{sec:freq_dist}, at the highest count rates (>2500 cts/s/PCU), the lightcurve becomes non-stationary and no discernible bicoherence pattern is observed. 

\subsection{Bicoherence and radio property correlation}

\cite{Muno2001} studied the radio and X-ray properties of the hard state of GRS~1915+105, and classified three states based on the radio properties. In the radio faint state, the radio flux (at 15.2 GHz) is below 20mJy. The radio bright (brighter than 20mJy at 15.2 GHz) observations can be classified into two states: radio plateau and radio steep. In the radio plateau state, the spectrum is optically thick, with a spectral index of $\alpha$ > -0.2 ($S_{\nu} \propto \nu^{\alpha}$), and has been associated with a steady, compact jet \citep{Dhawan2000}. In the radio steep state, the spectrum is optically thin, with a spectral index of $\alpha$ < -0.2. They also note that as the QPO frequency decreases, the radio emission becomes brighter and optically thick.

Following the classification of \cite{Muno2001}, \cite{Yan2013} found that the relation between the  LFQPO frequency and the hardness ratio formed two distinct branches (named `Branch 1' and `Branch 2'). It was noted by them that `Branch 1' corresponds to radio bright observations, while `Branch 2' consists of radio quiet observations. In `Branch 1', when the QPO frequency is below 2Hz, the source is in the radio plateau state. As the QPO frequency increases, the source evolves from the radio plateau to the radio steep conditions. 

\begin{figure*}
	\includegraphics[width=\textwidth]{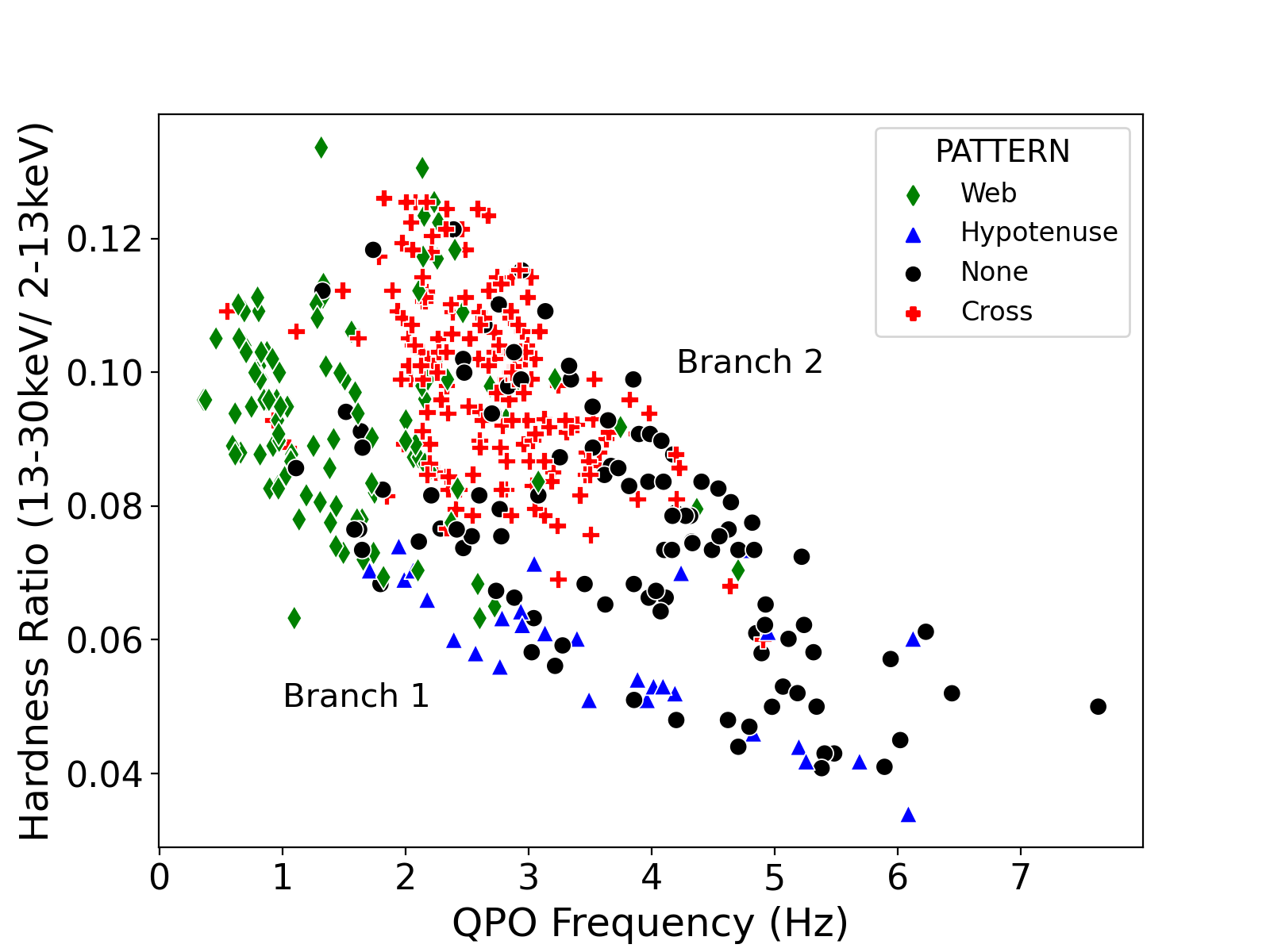}
    \caption{Hardness ratio of the observations plotted against the QPO frequency.`Branch 1' corresponds to radio bright observations, with observations where f$_{QPO}$ < 2Hz being in the plateau state and f$_{QPO}$ > 2Hz being radio steep. `Branch 2' consists of radio quiet observations. The coloured markers indicate the bicoherence pattern seen in each observation.}
    \label{fig:hardness_freq}
\end{figure*}

In our analysis, we find that the majority of the `Branch 2' observations show a `cross' pattern in the bicoherence. The `Branch 1' observations typically show a `web' pattern when the QPO frequency is below $\sim$2 Hz, and a `hypotenuse' pattern for QPO frequencies above $\sim$2 Hz (see Fig.~\ref{fig:hardness_freq}). In the observations that lie between branches 1 and 2 where the hardness ratio is greater than $\sim$ 0.07, the bicoherence patterns go from exhibiting a `web' pattern to `cross' pattern with an increase in QPO frequency. This is consistent with previous studies of the bicoherence of high inclination sources as discussed in Section~\ref{sec:freq_dist}.

When the source is in `Branch 2', the X-ray luminosity is low, and GRS~1915+105 is in a radio faint state, this corresponds to observations where the `cross' pattern is predominantly seen in the bicoherence, and the QPO frequency is above 2Hz. At frequencies greater than $\sim$3 Hz, the `web' pattern is seen on this branch. However, as outlined in Section~\ref{sec:freq_dist}, these observations are likely to be `cross' observation misidentified as `web' due to the presence of a broad sub-harmonic feature. 

When in `Branch 1', GRS~1915 is radio and X-ray bright. If the QPO frequency is below 2Hz, it is in a radio plateau state with optically thick radio emission, and the `web' pattern is predominantly seen in the bicoherence. However, as the QPO frequency increases above 2Hz, and the source evolves to a radio steep state with optically thin radio emission, the `hypotenuse' pattern is predominantly seen in the bicoherence. It is noteworthy that this QPO frequency of $\sim$ 2Hz, where the optical depth of the jet and the observed bicoherence pattern changes, coincides with the change in the QPO phase lag from positive to negative, and where no energy dependence of the QPO frequency is seen. Table~\ref{tab:properties} lists the properties of the observation when each bicoherence pattern is observed. 

\begin{table}
	\centering
	\caption{Table of the average X-ray and radio properties when each bicoherence pattern is observed. X-ray count rate is given in cts/s/PCU. See full text for details.}
	\label{tab:properties}
	\begin{tabular}{lcccr} 
		\hline
		Bicoherence & X-rays & Hardness & QPO  & Radio\\
		Pattern    & (countrate) & Ratio & Frequency & (optical depth) \\
		\hline
		Web & 1560 & Hard (0.09) & <2Hz & Bright (thick)\\
		Cross & 1380 & Hard (0.1) & 2-4 Hz  & Faint\\
		Hypotenuse & 2240 & Soft (0.06) & 2-4 Hz & Bright (thin)\\
		\hline
	\end{tabular}
\end{table}

\subsection{Bicoherence and average QPO phase lag correlation}

It has previously been established that the average phase lag at the fundamental of the QPOs (hereafter average phase lags) of GRS~1915+105 changes sign from a positive to negative at frequency of around 2 Hz \citep{Reig2000a, Pahari2013}. It has also been noted that there is a break slope of this correlation at $\sim$ 1.8 Hz \citep{Zhang2020}. Additionally, it was noted by these authors that the average QPO phase lags above 2Hz show a significantly larger scatter than would be expected from statistical fluctuations compared to those below 2Hz. As the observations with QPO frequency above 2Hz are known to exhibit two different bicoherence patterns, we investigate the possibility that the observations that show `hypotenuse' and `cross' patterns exhibit different phase lag behaviour. Here we only consider the phase lags at the QPO fundamental, and not at the (sub)harmonics.

\begin{figure}
	\includegraphics[width=\columnwidth]{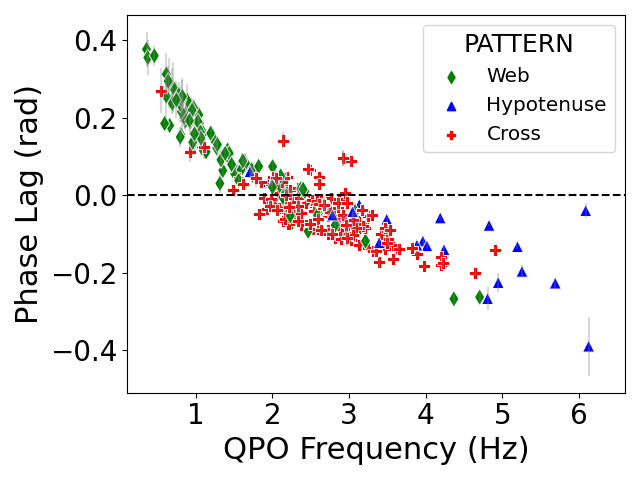}
    \caption{The lag- frequency spectrum of the observations. The horizontal line indicates zero phase lag. The lags are calculated between the 2-5.7 keV and 5.7-15 keV energy bands. Positive values indicate hard lags, and negative vales indicate soft lags. }
    \label{fig:lag_freq}
\end{figure}

\begin{figure}
    \centering
	\includegraphics[width=\columnwidth]{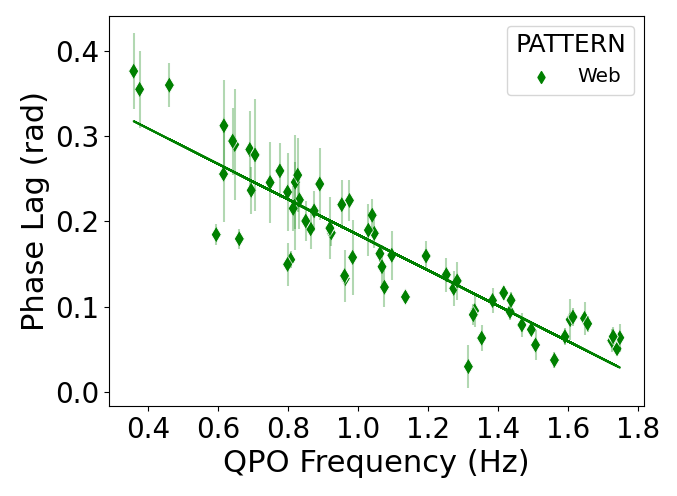}
    \newline
	\includegraphics[width=\columnwidth]{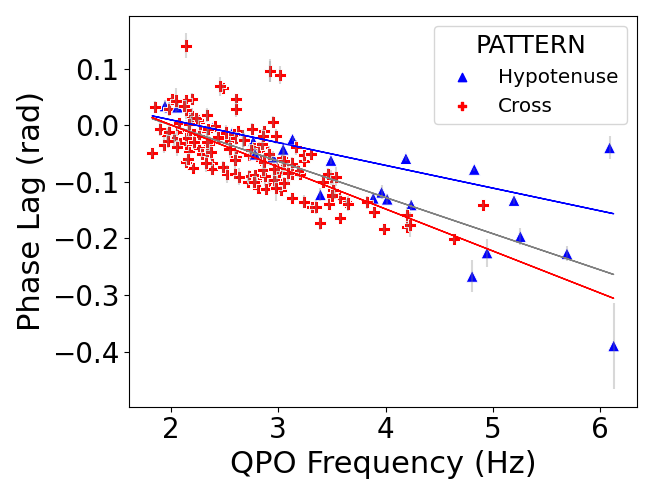}
    \caption{Top panel: The line of best fit for phase lags of observations that show a `web' pattern below 1.8Hz. Bottom panel: The line of best for the phase lags of observations that show `cross' and `hypotenuse' pattern separately, as well as when combined into a single population shown in grey.}
    \label{fig:lag_fit}   
\end{figure}

In Fig.~\ref{fig:lag_freq}, we plot the QPO average lags as a function of QPO frequency for each bicoherence pattern. It can be seen that the average phase lags go from a positive to a negative value at $\sim$ 2Hz, consistent with previous studies. First, as a consistency check, we fit the lag-frequency spectra of observations with QPO frequency below 1.8 Hz with a straight line, as shown in the top panel of Fig.~\ref{fig:lag_fit}. We find a slope of -0.21$\pm$0.02 for these observations, consistent with the value of the slope found by \cite{Zhang2020} below the break at 1.8Hz. This sample below $\sim$ 2Hz is dominated by observations that show the `web' pattern, and thus these observations were used for the fit. However, we confirm that the best fit value for the slope does not significantly change if the few observations that show the `cross' pattern are included.

We then fit the lag-frequency spectra of `hypotenuse' and `cross' pattern observations, and find slopes of -0.04$\pm$0.01 and -0.07$\pm$0.01 respectively.For these fits, we do not include the `web' pattern observations that have a QPO frequency greater than $\sim$2 Hz as they cannot be unambiguously classified as `web' or `cross' (see Section~\ref{sec:freq_dist}). This is shown in the lower panel of Fig.~\ref{fig:lag_fit}. With the `hypotenuse' and `cross' observations pooled, we find a slope of -0.06$\pm$0.01, which we note is higher than the slope of -0.10$\pm$0.01 found by \cite{Zhang2020}.

We conducted a parametric T-test to establish that the slopes of the `hypotenuse' and `cross' observations are different at a 95\% confidence level. However, as only a small number (26) of observations show the `hypotenuse' pattern, the best fit slope for this population may not be a reliable fit, especially considering the relatively large scatter in their phase lags at higher frequencies. Thus, to determine whether the two samples (i.e phase lags of `cross' and `hypotenuse' observations) are drawn from statistically different distributions, we performed a two dimensional KS test \citep{Fasano1987,Ness-Cohn2021} with the null hypothesis that both are drawn from the same underlying distribution. We find that with a p-value of p=0.002, the null hypothesis can be rejected.

We also note that the energy dependent phase lags in the plateau state and the radio-quiet states show complex behaviour, with the radio quiet states showing negative time lags that decrease with energy, the plateau states showing positive lags that increase with energy \citep{Pahari2013}. Furthermore, in addition to the change from positive to negative phase lags with QPO frequency, \cite{Pahari2013} also note a difference in the shape of the lag-frequency spectrum between radio quiet and plateau states. In the plateau state observations, the phase lags are roughly constant (especially at low frequencies) at some positive value. But during the radio-quiet state, the lags are negative (i.e. soft) at low frequency and show a dip-rise pattern, with a dip at the QPO frequency, above which the lag increases sharply. 

Additionally, in the radio steep observations that show the `hypotenuse' pattern, we find that the lag-frequency shows similar features to the plateau (`web') observations, rather than the dip-rise pattern seen from the radio quiet (`cross') observations. Thus it is likely that there are correlations between the phase-lag behaviour and the bicoherence pattern observed from each observation.


Several models have been proposed to explain the phase lag behaviour of GRS~1915+105 and other black hole X-ray binaries. \cite{Nobili2000} proposed that the change in sign of the phase lags at higher QPO frequencies was was due to a decrease of inner disc radius and an increase in the optical depth of the corona. \cite{Eijnden2016} consider the differential precession of the inner accretion flow in order to explain the change in the average phase lag, as well as the energy dependence of the QPO centroid frequency. In this scenario, at high QPO frequencies (> 2 Hz) the inner half has a harder spectrum than the outer half, and the opposite is true at higher QPO frequencies. \cite{Zhang2017} also consider the possibility that the hard and soft lags originate from different mechanisms, with the reflection component contributing significantly to the average phase lags at high QPO frequencies. 

\cite{Karpouzas2021a} consider a Comptonisation model and find that a corona that varies in size, along with a switch in the efficiency of heating of the seed photon source by photons that have been Compton up-scattered in the corona, is able to explain the observed lag-frequency spectrum, and the frequency dependent lag-energy and rms-energy and spectra from GRS~1915+105. This model is able to explain the phase-lag properties of GRS~1915+105, as well as cases where the lag-energy spectra is complex (such as energy dependence of the lags and rms amplitude of the type-B QPO in MAXI~J1348−630, \cite{Garcia2021}). However, due to the complexity and current lack of understanding of the differences between the phase lags of high and low inclination sources \citep{VandenEijnden2017}, further work is needed to reproduce the source inclination dependent behaviour of QPO phase lags.

The bicoherence properties of QPOs are also known to exhibit source inclination dependent behaviour \citep{Arur2020}. Thus, a more extensive examination of the phase lags as they relate to the radio and bicoherence properties, will be useful in determining how the interplay between the jet, disc and corona contributes to the X-ray variability. As the phase-lags are known to be energy dependent, a detailed analysis of the relation between the phase lag, bicoherence and the energy dependent bispectral properties (the biphase) will be presented in future work.

\section{Discussion}

In this paper, we have presented the results of a systematic analysis of the bicoherence of type C QPOs from GRS~1915+105. We find that below a QPO frequency of $\sim$ 2Hz, the `web' pattern dominates, and these observations coincide with optically thick radio emission. When the QPO frequency is above $\sim$ 2Hz, either a `cross' or `hypotenuse' pattern is seen. The `cross' pattern coincides with spectrally hard X-ray observations and optically thin radio emission, while the `hypotenuse' pattern coincides with observations that are spectrally soft in the X-rays and are radio faint. In this section, we will discuss these results and interpret the observed correlations between the bicoherence, X-ray and radio properties.

\subsection{Interpreting the correlations between the bicoherence, X-ray and radio properties}
\label{sec:interpretation}

GRS~1915 is considered a high inclination (edge-on) source based on the inclination of its relativistic jets \citep{Mirabel1994} and the assumption that any misalignment between the binary orbit and the jet is small. Additionally, GRS~1915+105 was also the first source to show evidence of strong disc winds from absorption line studies \citep{Lee2002, Neilsen2009}. Thus, based on previous studies of the bicoherence \citep{Arur2020}, GRS~1915 would be expected to show a gradual change from the `web' pattern to the `cross' pattern as the QPO evolves to higher frequencies. Indeed, as expected, the `web' pattern is observed at low (<2 Hz) frequencies, with the `cross' pattern seen at higher (>2 Hz) frequencies. However, at the highest count rates while still being non-stationary, the `hypotenuse' pattern is seen. Apart from the unique case of GRS~1915+105, this pattern is only observed from low inclination (face-on) sources.

In order to explain the behaviour of the bicoherence and its correlation with the X-ray and radio properties in GRS~1915+105, we consider the truncated disc model \citep{Ichimaru1977,Esin1997}, where at low accretion rates, the inner edge of the accretion disc is truncated at a radius greater than the innermost stable circular orbit (ISCO). We interpret our results in the context of the Lense-Thirring precession model \citep{Stella1998, Ingram2009}, where the QPO is associated with X-ray modulation arising from the precession of a geometrically thick, toroidal accretion flow that is located inside the truncated accretion disk. 

For the radio properties, we consider a conical compact jet which has a Synchrotron spectrum at each height in the jet, with a characteristic break frequency. At frequencies above the break frequency, the radio spectrum is optically thin, synchrotron spectrum due to electrons accelerated by Fermi acceleration. Below this frequency, the spectrum is optically thick due to synchrotron self-absorption (see eg. \cite{Rybicki1979}). The break frequency decreases linearly with distance from the black hole along the jet, giving rise to a flat radio spectrum \citep{Blandford1979}.

First we consider the case of when the `cross' pattern is seen, where the source is hard and is faint in the radio. At this accretion rate (an average of $\sim$ 1300 counts/sec/PCU, see Table~\ref{tab:properties}), the inner accretion region can be considered to be geometrically thin with high optical depth ($\tau \gg $ 1) \citep{Shakura1973}. Due to the high optical depth, small increases in the mass accretion rate do not lead to a corresponding increase in the X-ray luminosity. As a result, no amplitude modulation is present from fluctuations in the mass accretion rate and the diagonal structure indicating phase coupling with the low frequency broadband noise is not observed. However, the precession of the inner accretion region still modulates the high frequency variability as a function of the QPO phase due to a variation in the optical depth along the line of sight \citep{Arur2020}. Thus, the resultant cross pattern is observed. The presence of jets has been attributed to the presence of poloidal magnetic fields that thread the accretion disc, with the strongest jets present in a geometrically thick disc \citep{Meier2001}, with observations of radio emission being quenched as the disc becomes geometrically thin \citep{Tananbaum1972a,Russell2019}. Thus in this thin disc regime, the geometrically thin disc quenches the jets, explaining the radio faintness when the `cross' pattern is seen.

The observations where the `web' pattern is seen to occur when the source is hard, the QPO frequency is low (<2Hz) and where the radio emission is bright and optically thick. To determine the if statistically significant difference is present between the mean QPO frequency of the `cross' and `web' patterns (at 2.67Hz and 1.56Hz respectively), we conducted a T-test and found that the mean of the two sets of observations are different at a confidence level $>$5$\sigma$. Here, the low QPO frequency when the `web' pattern is observed is interpreted as the truncation of the inner disc at a larger radius. In these observations, the formation of a steady compact jet would require a geometrically thick accretion disc to support the poloidal magnetic fields. In such cases, if the disc structures changes and has an increased scale height due to radiation pressure, this would lead to a lower optical depth for the same mass accretion rate. It has been shown that in certain accretion regimes (L/L$_{Edd}$ = 0.1-1), the thickness of the disc increases with the angular momentum distribution still being Keperlian \citep{Abramowicz1988}. For the observations we consider (L/L$_{Edd} \sim$ 0.6), it is plausible that the accretion disc is in this regime.

In the case where fluctuations in the mass accretion rate propagate inwards and modulate the X-ray luminosity \citep{Lyubarskii1997,Arevalo2006}, the resulting amplitude modulation gives rise to the diagonal feature on the lower left side of the bicoherence plots. It has been shown that the optical depth in accretion discs can vary substantially depending on the accretion rate and the ($\alpha$) viscosity parameter \citep{Sadowski2011}. It has been estimated \citep{Arur2019} that at moderate optical depths ($\tau \sim$ 3), the high frequency ($f >$ 10Hz) variations are smeared out as the scattering timescales in optically thick regions are longer than the timescale of the variations, while lower frequency variations are preserved. For the observations we consider, as the QPO frequency is low compared to scattering timescales, it is plausible that the modulation required to produce the observed bicoherence will be visible. 

\begin{figure}
	\includegraphics[width=\columnwidth]{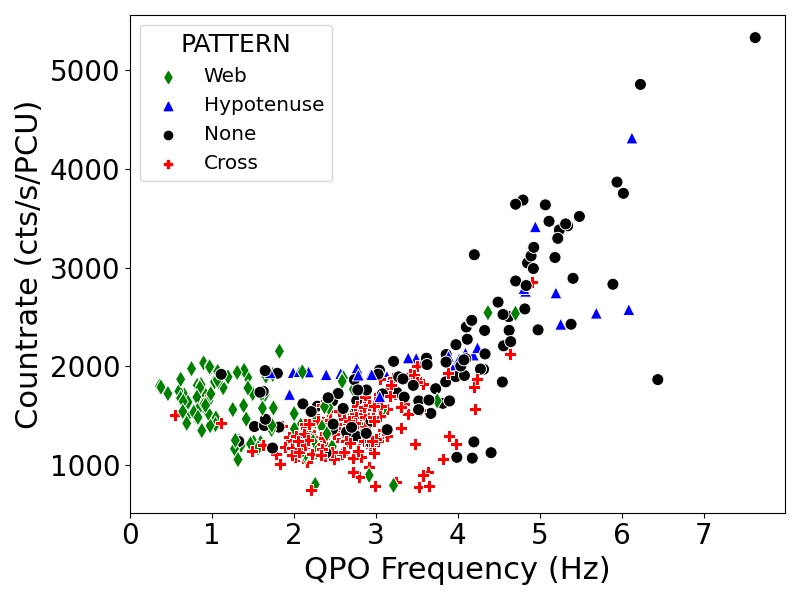}
    \caption{The countrate of the observations as a function of QPO frequency. The coloured markers indicate the bicoherence pattern seen during each observation.}
    \label{fig:freq_count}
\end{figure}

Additionally, the precession of an optically thick Comptonising region would modulate the high frequency variations observed from the innermost region with the QPO phase when viewed at high inclination (edge-on). This gives rise the vertical and diagonal features in the bicoherence. Overall, this scenario results in the `web' pattern of the bicoherence and is consistent with the observations of optically thick radio emission when this pattern is seen. 

The observations where the `hypotenuse' pattern is seen coincides with bright ($\geq$ 1700 cts/s/PCU) X-ray observations, along with optically thin radio emission. At these high accretion rates, the accretion disc is in a similar radiation pressure dominated configuration to when the `web' pattern is observed, as described above. In such an optically thin region, mass accretion rate fluctuations result in amplitude modulation of the X-rays.

As the inner edge of the disc moves inwards (indicated by the increase in QPO frequency), and the spectrum becomes increasingly dominated by the accretion disc (as indicated by the decrease in the hardness ratio), a high influx of soft disc photons is experienced by the inner accretion region. If a significant fraction of the energy from the disc photons is intercepted by the non-thermal electrons in the jet, this could lead to a cooling of the jet electrons through inverse Compton scattering. This cooling would then result in weakening the Synchrotron self-absorption, leading to an optically thin jet. This is consistent with the optically thin radio emission seen when the `hypotenuse' pattern is observed. In this case, as the high frequency variations arise from an optically thin region, these variations would not be modulated on the QPO phase, and no vertical or horizontal features would be seen in the bicoherence. Thus, only the diagonal feature is observed, resulting in the hypotenuse pattern in the bicoherence.

Inverse Compton scattering of soft disc photons by electrons in the jet has been invoked to explain frequency dependent time lag spectra in the hard state of black hole X-ray binaries \citep{Reig2003}, and this model has been used to reproduce the energy spectra from radio to X-rays of XTE~J1118+480 and Cygnus~X-1 \citep{Giannios2005}

The unique behaviour of GRS~1915 (i.e the `hypotenuse' pattern being observed in a high inclination source) could be explained using this scenario, as at the highest X-ray luminosities where a discernible pattern seen, we observe the `hypotenuse' pattern, and GRS~1915+105 is known to be the only lack hole binary in our Galaxy that spends significant periods at very high luminosities. Indeed, it has been the only source to reach X-ray luminosities high enough to trigger the unique, limit cycle like behaviour seen from it, possibly arising from a radiation pressure instability \citep{Done2004,Belloni1997}. However, as the `hypotenuse' pattern is also observed at lower luminosities, it remains to be seen if this scenario can fully explain the observed behaviour. We also note that as only a small fraction of the observations show a `hypotenuse' pattern, more observations that show this pattern are required for a detailed examination of its dependence on the X-ray luminosity and the properties of the accretion disc in this state. 

Above, we presented a scenario in the context of the Lense-Thirring precession model to explain the observed correlations between the radio properties and the bicoherence in GRS~1915+105. However various details, such as the conditions under which the accretion disk starts to become radiation pressure supported, are unclear. Previous studies of type-C QPOs have shown that in high inclination (edge-on) sources, the bicoherence shows a gradual change from `web' to `cross' as the QPO frequency increases and hardness ratio decreases. The increase in QPO frequency is used as proxy for the inner edge of the disc moving inwards as the mass accretion rate increases. However, in the case of GRS~1915+105, both the `web' and `cross' patterns have significant overlap in their hardness ratio and X-ray countrate. Additionally, a number of observations at high (>2 Hz) frequency that show the `cross' pattern have countrates below that of low frequency (<2 Hz) observations that show the `web' pattern (see Fig.~\ref{fig:freq_count}), which cannot be explained through an increase in the mass accretion rate. On the other hand, if the disc is in the regime where radiation pressure starts to become an important factor, the relatively small increase in the accretion rate could be lead to an increase in the scale height of the disk and thus result in a radio bright state. All together, this indicates that a complex interplay between the disc-corona-jet system is required to fully explain the observed behaviour, which is beyond the scope of the illustrative scenario presented in this paper.

Recent analysis by \citep{Mendez2022} of observations from GRS~1915+105 found that the radio flux from the jet and the flux of the iron line are strongly anti-correlated with the electron temperature of the X-ray corona and the amplitude of the high frequency(60-80Hz) variability component coming from the innermost part of the accretion flow. These results are interpreted such that the energy powering these systems is directed either towards jet acceleration or towards heating the corona, leading to correlations with the X-ray flux. This also provide additional evidence that interactions between the accretion disc, jet and corona is required to explain the correlations observed with the bicoherence behaviour presented in this paper (cf. Figure 1 of \citet{Mendez2022} and Figure~\ref{fig:hardness_freq} in this work).

We emphasize that currently there is a lack of detailed models that aim to explain the bispectral behaviour of black hole X-ray binaries. Thus, the above interpretation is presented as an outline to motivate analytic efforts to explain the observed bispectral behaviour of QPOs from GRS~1915+105. We also note that direct comparisons between the observed bicoherence with those from numerical models require simulated light curves that have a combination of sufficient length and time resolution to qualitatively evaluate different models for the QPO mechanism. While such an approach is presently computationally intensive, advances in computational modelling could enable such comparisons in the future. 

\subsection{Jet energetics}

As discussed in Section~\ref{sec:interpretation}, we observe that there is a strong correlation between the bicoherence pattern and the radio properties in GRS 1915+105.  This is not surprising, given that there was already a well-known correlation between the radio properties and the QPO frequencies.  It is thus natural to investigate whether the accretion rate and geometry needed to explain the bicoherence properties can also lead to an explanation for the jet properties.  In particular, we find that it is plausible that the energy density of the disc photons is sufficient to cool the jet effectively enough, and reduce the strength of the synchrotron self absorption during the `hypotenuse' states, leading to the correlation between those states and their steep spectrum radio emission.

The steep spectrum jets are usually presumed to arise from discrete ejection events. If this is the case, it would be surprising for the accretion disc's spectrum and power spectrum to look so similar to scenarios in which the jet is a conical compact jet.  We thus consider an alternative scenario in which the jet's basic structure is similar in the radio-steep states, but is affected by external photons Compton cooling the jet, and test the plausibility of that scenario.  This is analogous to the ``blazar sequence'' for supermassive black holes, in which the blazars with strong thermal emission from the accretion flow show ``redder" spectra, with weaker synchrotron luminosities relative to their Compton luminosities \citep{Fossati1998,Ghisellini2008}.

Here, we consider a conical jet with a base height of 10$^9$ cm (optical and infrared time delays with respect to X-rays from black hole binaries are of the order of 0.1 seconds, see e.g \cite{Casella2010,Gandhi2017}) and an opening angle of 10 degrees. Radio imaging of the jets of GRS~1915+105 indicate that the jets are optically thick at 15GHz out to $\sim$ 50 AU in the plateau state \citep{Dhawan2000, Klein-Wolt2002a}.

To test whether cooling of the jet can happen via disc seed photons, we wish to establish that (1) the disc seed photons inject more power into the jet than its intrinsic synchrotron power and (2) that the disc seed photon power also exceeds the synchrotron self-Compton power.  By considering the solid angle subtended by the jet from the surface of an isotropically radiating disc, we find that approximately $\sim$ 42\% of the disc photons are intercepted by the jet, so, unless the jet's intrinsic synchrotron power is comparable to the intrinsic disc power, or the jet is strongly beamed away from the disc, the seed photons for Comptonization in the jet will be dominated by the disc photons.  The former point can be established in a straightforward manner given that the jet is steep spectrum even in the radio band in the states of interest.  The brightest of the ``radio-steep'' observations in \citep{Muno2001} show $F_R\approx 10^{-14}$ erg/sec/cm$^2$ near 10 GHz, while the X-ray fluxes are typically $\sim10^7$ times higher than this value.  For the synchrotron emission with spectral index of $-0.7$ then to have comparable power to the X-rays from the disc within the jet, it would need to continue out far into the gamma-ray band without a cooling break.  We find this to be rather unlikely.

We emphasize that the estimated value for the fraction of photons that are intercepted is sensitive to the base height of the jet, which is currently not well determined, is thus not a robust estimate. In future, multi-wavelength jet variability and eclipse mapping have the potential to provide an estimate of jet properties including the base height \citep{Tetarenko2019,Maccarone2020}; some of this work has already been undertaken for variability studies \citep{Tetarenko2019,Tetarenko2021}. 

The energy density of the disc photons can be estimated to be $U_{ph} = L / 4\pi d^2 c$ where L is the X-ray luminosity and d is the distance from the disc to the base of the jet. During this bright stage, the countrate is $\sim$2000 counts/sec, which corresponds to a luminosity of 3$\times$ 10$^{38}$ ergs/sec, using values of $N_H = 5 \times 10^{22} {\rm cm}^{-2}$ \citep{Lee2002}, $\Gamma$ = 1.8 and a distance of d = 8.6kpc \citep{Reid2014}.

The magnetic energy density $U_B$ is given by $B^2/8 \pi$ \citep{Rybicki1979} , where $B$ is the magnetic field at the base of the jet and is assumed to be $\approx 10^5$ Gauss, based on previous estimates from other sources ($10^5 - 10^7$G in Cyg~X-1 \citep{DelSanto2013}, $\sim 5 \times 10^4$G in XTE J1550-564 \citep{Chaty2011} $\sim$ 1.5 $\times 10^4$G in GX~339-04 \citep{Gandhi2011}).We note that we do not take into account the beaming of the jet, which could affect the injection of soft disc photons, in this calculation, as such detailed modelling is beyond the scope of this work. Assuming that approximately $\sim$40\% of the disc photon energy density is intercepted by the disc, and that this is the dominant source of seed photons, as suggested above, the ratio of photon to magnetic energy density in the base of the jet $U_{ph} / U_B \sim 0.85 $. This indicates that these energy densities are comparable, and that the cooling of the jet by the disc photons could be an important factor, but that this only matters at the brighter X-ray luminosities. This is consistent with the fact that the `hypotenuse' pattern is only visible during observations where the X-ray count rate is high.

\section{Conclusions}

We have explored the behaviour of the bicoherence of a number of QPOs from GRS~1915+105 to investigate the coupling between the phases of different Fourier frequencies. GRS~1915+105 is unique in its bicoherence behaviour as all three known bicoherence patterns displayed by type-C QPOs are observed. This is in contrast with other black hole binaries which only show two of the three patterns.  We find that the nature of the bicoherence seen correlates with the radio properties of the source at the time of the observations, the QPO frequency as well as the hardness ratio in the X-rays as summarised below:

\begin{enumerate}
    \item When the QPO frequency is below $\sim$2 Hz, and the source is in the radio plateau state with optically thick radio emission, the `web' pattern is predominantly seen.
    \item When the QPO frequency is above $\sim$2 Hz and the source is in the radio steep state with optically thin radio emission, the `hypotenuse' pattern is predominantly seen. During these observations, the source is bright and soft in the X-rays.
    \item When the QPO frequency is above $\sim$2 Hz and the source is radio faint, the `cross' pattern is predominantly seen. During these observations, the source is hard in the X-rays.
\end{enumerate}

We interpret these results in the context of the Lense-Thirring precession model and find that this unique bicoherence behaviour of GRS~1915+105, along with its correlation with the radio properties can be explained if a significant fraction of soft photons from the disc are intercepted by the jet. This could lead to the cooling of the jet electrons, making the jet optically thin when the source is bright in the X-rays. This behaviour is unique to GRS~1915 due to the high X-ray luminosities it reaches.

We find the frequency dependent phase lag behaviour of the `cross' and `hypotenuse' patterns to be consistent with being drawn from different populations. A more detailed investigation of energy dependent and frequency dependent phase lag and bispectral behavior will be the subject of future work.

Finally, we have shown that the use of higher order timing analysis techniques like the bicoherence has the ability to provide new insights into the accretion process of black hole X-ray binaries, distinguishing different physical processes despite them producing similar features in the power spectrum. However, there still remain open questions on the exact nature of those processes, and these results highlight the need for detailed and sophisticated modelling of the accretion flow and jet of GRS~1915+105.

\section*{Acknowledgements}
The authors would like to thank Liang Zhang for providing a list of observations and QPO parameters. We also thank the anonymous referee for comments that improved the clarity and content of the paper. This research has made use of data and software provided by the High Energy Astrophysics Science Archive Research Center (HEASARC), which is a service of the Astrophysics Science Division at NASA/GSFC, as well as NASA's Astrophysics Data System Bibliographic Services.

\section*{Data Availability}

No new observational data are presented in this paper. The observations used for the analysis presented in this paper are listed in the supplementary data table available in the online version and publicly available from the NASA HEASARC.



\bibliographystyle{mnras}
\bibliography{GRS1915_new} 




\bsp	
\label{lastpage}
\end{document}